\documentstyle[twocolumn]{mn}
\input{epsf}

\newcommand{\bez}{\begin{eqnarray*}}
\newcommand{\eez}{\end{eqnarray*}}
\newcommand{\be}{\begin{equation}}
\newcommand{\ee}{\end{equation}}
\newcommand{\beq}{\begin{eqnarray}}
\newcommand{\eeq}{\end{eqnarray}}
\newcommand{\bc}{\begin{center}}
\newcommand{\ec}{\end{center}}

\newbox\grsign \setbox\grsign=\hbox{$>$} \newdimen\grdimen \grdimen=\ht\grsign
\newbox\simlessbox \newbox\simgreatbox \newbox\simpropbox
\setbox\simgreatbox=\hbox{\raise.5ex\hbox{$>$}\llap
     {\lower.5ex\hbox{$\sim$}}}\ht1=\grdimen\dp1=0pt
\setbox\simlessbox=\hbox{\raise.5ex\hbox{$<$}\llap
     {\lower.5ex\hbox{$\sim$}}}\ht2=\grdimen\dp2=0pt
\setbox\simpropbox=\hbox{\raise.5ex\hbox{$\propto$}\llap
     {\lower.5ex\hbox{$\sim$}}}\ht2=\grdimen\dp2=0pt
\def\simgreat{\mathrel{\copy\simgreatbox}}

\begin{document}

\title[Super-Eddington accretion disc around a Kerr black hole]
{Super-Eddington accretion disc around a Kerr black hole}

\author[A.~M. Beloborodov]
{\parbox[]{6.8in} {A.~M. Beloborodov} \\
Astro Space Centre of P.N. Lebedev Physical Institute, 
84/32 Profsoyuznaya Street, Moscow 117810, Russia}

\date{Accepted, Received}

\maketitle


\begin{abstract}

We calculate the structure of accretion disc around a rapidly rotating 
black hole with a super-Eddington accretion rate.
The luminosity and height of the disc are reduced by the 
advection effect. In the case of a large viscosity parameter, $\alpha > 0.03$, 
the accretion flow strongly deviates from thermodynamic 
equilibrium and overheats in the central region.
With increasing accretion rate, the flow temperature steeply increases,
reaches a maximum, and then falls off. The maximum is achieved in the 
advection dominated regime of accretion. The maximum 
temperature in the disc around a massive black hole, $M=10^8M_\odot$,
with $\alpha=0.3$ is of order $3\times 10^8$ K.
Discs with large accretion rates can emit X-rays 
in quasars as well as in galactic black hole candidates. 

\end{abstract}



\def\ep{\varepsilon}
\def\gga{\gamma-\gamma}
\def\bbeta{\mbox{\boldmath $\beta$}}
\def\bv{\mbox{\boldmath $v$}}
\def\Md{\dot{M}}
\def\md{\dot{m}}

\section{Introduction}

Black hole accretion discs are believed to be the mechanism of energy release
in active galactic nuclei and in some X-ray binaries. 
Their luminosity is constrained by the Eddington limit, 
$L_E=4\pi GMm_pc/\sigma_T$ ($M$ is the black hole mass), and  
the accretion rate, $\Md$, is usually expressed in the Eddington units,
$\md=\Md c^2/L_E$. The luminosity becomes equal to the Eddington limit
at $\md_{cr}=\eta^{-1}$, where $\eta$ is the radiative efficiency of the disc.
Accretion rates of order $\md_{cr}$ are likely to feed the huge energy 
release in quasars, and $\dot{m}\sim \dot{m}_{cr}$ is implied in many models 
of the soft X-ray excess observed in quasar spectra 
(e.g. Czerny \& Elvis 1987; D\"{o}rrer et al. 1996; Szuszkiewicz, Malkan
\& Abramowicz 1996). 

According to the standard model (Shakura 1972; Shakura \& Sunyaev 1973),
the effective temperature of the disc surface scales as $T^s_{eff}\propto 
(\md/M)^{1/4}$.  A special feature of discs with $\dot{m}\sim\dot{m}_{cr}$
is that their temperature can be much greater than $T_{eff}^s$ provided
the viscosity parameter $\alpha$ is large enough (Shakura \& Sunyaev 1973). 
In the inner region of such a disc, the inflow time-scale is shorter than 
the time-scale for relaxation to thermodynamic equilibrium and the accretion 
flow is overheated. The overheating might be strong 
enough for a transition from the standard radiation 
dominated disc to the hot two-temperature ion-pressure supported regime of
accretion proposed by Shapiro, Lightman \& Eardley (1976). This transition
has been discussed in recent works (Liang \& Wandel 1991; Artemova et al.
1996; Bj\"{o}rnsson et al. 1996).

The standard model can be a good approximation only if 
$\dot{m} < \dot{m}_{cr}$. When the accretion rate approaches 
$\dot{m}_{cr}$ the main assumption of the model, that
the viscously released energy is radiated away locally, becomes unadequate. 
Inside some radius $r_t$, the produced radiation
is trapped by the flow and advected eventually to the black hole instead
of being radiated away. This radius can be estimated by taking the internal 
energy density, $\varepsilon$, and the density, $\rho$, of the disc from 
Shakura \& Sunyaev (1973). Then 
one finds that the radial flux of internal energy  $=\dot{M}\varepsilon/\rho$ 
exceeds the total flux of radiation emitted between $r$ and $2r$ at 
\begin{equation}
r_t\approx \md r_g \sqrt{1-\left(\frac{3r_g}{r_t}\right)^{1/2}},
\end{equation}
where $r_g=2GM/c^2$ is the gravitational radius.
For $\dot{m}=\md_{cr}=12$ equation (1)
yields $r_t\approx 7.1 r_g$. A substantial portion of the 
released energy must be swallowed by the black hole even when $\md<\md_{cr}$.
On the other hand, the relative height of the Shakura-Sunyaev disc, $H/r$, 
equals $\md/27$ at the maximum.
So, advection becomes essential before the accretion flow
becomes quasi-spherical, and this effect can be 
investigated in an extended version of the standard model retaining the 
vertically integrated approximation. The corresponding set of equations was 
proposed by Paczy{\'n}ski \& Bisnovatyi-Kogan (1981). 
Numerical solution of these equations gives adequate description for
the inner transonic edge of the accretion flow and shows that the disc
remains relatively thin at moderately super-Eddington accretion rates
(Abramowicz et al. 1988; Chen \& Taam 1993). 
This made natural the extension of the standard model to the super-Eddington 
advection dominated regime, called "slim" accretion disc.

All the models of the super-Eddington slim disc employed the 
pseudo-Newtonian approximation to the black 
hole gravitational field (Paczy\'{n}ski \& Wiita 1980) which is good only in 
the case of Schwarzschild black hole. 
Recently, the equations of relativistic slim disc has been derived by Lasota 
(1994, see also Abramowicz et al. 1996) and applied to another type of 
advection dominated accretion flow, optically thin and hot.

In the present paper, we solve the relativistic equations of the
super-Eddington disc. In the case of a rapidly rotating black 
hole relativistic effects become especially important. The sonic radius 
is typically inside the ergosphere, close to the black hole horizon.
The released power steeply increases when the hole spin 
approaches its extreme value, $a_*=Jc/M^2G\rightarrow 1$ ($J$ is the  
angular momentum of the black hole). In this paper, we calculate the disc
structure for the case $a_*=0.998$ and compare it with the case of 
non-rotating black hole $a_*=0$.

We pay particular attention to the case of a large viscosity parameter, 
$\alpha$. In the previously investigated pseudo-Newtonian models, the 
super-Eddington disc was assumed to be in thermodynamic equilibrium which 
is a good approximation in the case of small $\alpha$. We find that a 
significant deviation from the equilibrium occurs when $\alpha\simgreat 0.03$. 
Then the flow overheats in the central region. We find that the temperature 
of the flow steeply rises at $\md>\md_{cr}$ and reaches a maximum in the 
advection dominated regime of accretion. The maximum 
temperature is especially high in the case of a rapidly rotating black hole.

In next Section, we write down the equations of the relativistic advective 
disc. The equations describe a radiation dominated flow and 
neglect the thermal pressure of the accreting plasma. 
Their solution yields the radiation density, plasma 
density, and height of the disc within the equipartition radius of
the standard model, $r_{eq}$, at which the radiation pressure equals the 
plasma pressure. The transition from the standard model 
to the advective accretion occurs at $r_t\ll r_{eq}$.
We describe the numerical method, present the solution, 
and discuss the structure of the disc in Section 3.
Then, in Section 4, we calculate the temperature of the accreting plasma.
Finally, we check that inside $r_{eq}$ our radiation-dominated models never 
approach the ion-pressure supported regime of accretion.
The results are summarised in Section 5.

\section{Relativistic disc equations}

In the description of the gas flow around a Kerr black hole 
we use the Boyer-Lindquist coordinates $x^i = (t, r, \theta, \varphi)$.
The metric tensor $g_{ij}$ of the Kerr space-time is given, 
e.g., by Misner, Thorne \& Wheeler (1973). 
The disc is assumed to lie at the equatorial plane of the Kerr geometry. 
The four-velocity of the accreting gas has components 
$u^i =(u^t, u^r, u^{\theta}, u^{\varphi})$ 
in the Boyer-Lindquist coordinates, and $u^\theta=0$. 
The angular velocity of the gas rotation equals $\Omega = u^\varphi/u^t$
and Lorentz factor measured in the frame of local observers with zero
angular momentum equals $\gamma = u^t(-g^{tt})^{-1/2}$. 

To obtain a simple model of the relativistic disc with advection we 
employ the vertically integrated (slim) approximation. 
The slim disc equations are written for the following
thermodynamic quantities:
surface rest mass density $\Sigma$, surface energy density $U$ 
(which includes both the rest mass energy $\Sigma c^2$ and the internal energy 
$\Pi$), and the vertically integrated pressure $P$. 
The dimensionless specific enthalpy is defined as $\mu=(U+P)/\Sigma c^2$. 
The equation of state in the radiation dominated disc is $\Pi\approx 3P$.
$F^+$ denotes the surface rate of viscous heating, and 
$F^-$ denotes the radiation flux radiated from both faces of the disc.
All the thermodynamic
quantities and both fluxes $F^-, F^+$ are measured in the comoving frame.

The main equations express
the conservation laws for barion number, energy, angular momentum,
and radial momentum. We include in the equations the inertial mass
associated with internal energy accumulated in the flow 
(Beloborodov, Abramowicz \& Novikov 1997).
\medskip

\noindent
{\it i) Barion conservation}

\begin{equation}
  2\pi r c u^r \Sigma =-\dot{M}.
\end{equation}

\noindent
{\it ii) Conservation of angular momentum}

\begin{equation}
  \frac{d}{dr}\left[\mu\left(\frac{\dot{M}u_\varphi}{2\pi}
  +2\nu\Sigma r \sigma_{~\varphi}^r\right)\right]=\frac{F^-}{c^2}\;ru_\varphi,
\end{equation}
where $\nu$ is the kinematic viscosity, and $\sigma_{~\varphi}^r$ is the 
shear, 
\begin{eqnarray}
  \sigma_{~\varphi}^r= \frac{1}{2}\;g^{rr}g_{\varphi\varphi}\sqrt{-g^{tt}}
  \gamma^3\frac{d\Omega}{dr}. \nonumber
\end{eqnarray}

\noindent
{\it iii) First law of thermodynamics}

\begin{equation}
   F^+-F^-=cu^r\left(\frac{d\Pi}{dr}
   -\xi \frac{\Pi+P}{\Sigma}\frac{d\Sigma}{dr}\right),
\end{equation}
where $\xi\approx 1$ is a numerical factor accounting for non-homogeneity
of the disc in the vertical direction, hereafter we set $\xi=1$.
The surface heating rate is given by
\begin{eqnarray}
\nonumber
  F^+=2\nu\Sigma \mu\; \sigma^2 c^2, \qquad
  \sigma^2=\frac{1}{2} g^{rr}g_{\varphi\varphi}\left(-g^{tt}\right)
  \gamma^4\left(\frac{d\Omega}{dr}\right)^2.  
\end{eqnarray}

\noindent
{\it iv) Radial Euler equation}

\begin{eqnarray}
\nonumber
   \frac{1}{2}\;\frac{d}{dr}\left(u_ru^r\right)=
   -\frac{1}{2}\;\frac{\partial g_{\varphi\varphi}}{\partial r}
   g^{tt}\gamma^2
   \left(\Omega-\Omega_K^+\right)\left(\Omega-\Omega_K^-\right)\\
  -\frac{1}{c^2\Sigma \mu}\frac{dP}{dr}-\frac{F^+u_r}{c^3\Sigma \mu},
\end{eqnarray}
where $\Omega_K^\pm$ are the Keplerian angular velocities,
\begin{eqnarray}
\nonumber
\Omega_K^\pm=\pm\frac{c}{r(2r/r_g)^{1/2}\pm a}.
\end{eqnarray}

The set of disc structure equations  becomes closed when the
viscosity, $\nu$, and radiative cooling, $F^-$, are specified. The standard 
$\alpha$-prescription for viscosity is $\nu=\alpha c_sH$, where $\alpha$ is
a constant, $c_s=c(P/U)^{1/2}$ is the isothermal sound speed. 
The half-thickness of the disc, $H$, should be
estimated from the vertical balance condition. 
\medskip

\noindent
{\it v) Vertical balance}

Near the black hole, relativistic effects become important and the 
tidal force contracting the disc in vertical direction depends on $\Omega$.
At $\Omega=\Omega^+_K$, the vertical tidal acceleration in the comoving 
tetrade equals (e.g. Riffert \& Herold 1995)
\begin{equation}
  a_{(\theta)}=\frac{zr_g(r^2-a_*r_g\sqrt{2r_gr}+0.75a_*^2r_g^2)}
  {r^3(2r^2-3r_gr+a_*r_g\sqrt{2r_gr})}=\frac{zr_g}{2r^3}\;J(a_*,r),
\end{equation}
where $z=\sqrt{g_{\theta\theta}}\;(\theta-\pi/2)$ is the height in the disc,
$J$ is a relativistic correction factor becoming unity at $r\gg r_g$.
Then the typical half-thickness of the disc can be estimated as
\begin{equation}
   H^2=\frac{P}{U}\;\frac{2r^3}{r_gJ}.
\end{equation}
More accurate expression accounts for a deviation $\Delta\Omega=
\Omega-\Omega_K^+$ of the gas rotation from Keplerian (Abramowicz, Lanza \&
Percival 1997). We will consider only the disc region outside the sonic
radius where the correction  to estimation (7) connected with $\Delta\Omega$ 
does not exceed several percent, and hereafter we use this estimation.
\medskip

\noindent
{\it vi) The radiative losses}

The time-scale for photon diffusion from the 
interior of the disc to its surface equals $t_D=H\tau_0/c$ where 
$\tau_0=\sigma_T\Sigma/2m_p$.
This is a typical leaking time for the radiation trapped inside the disc,
and the radiative losses can be written as 
\begin{equation}
  F^-=\chi\; \frac{m_p c\;\Pi}{\sigma_T\Sigma H},
\end{equation}
where $\chi\sim 1$ is a numerical factor. In the standard
model with the vertical structure governed by the radiation diffusion
this factor equals $2/\sqrt{3}$.
In the advection dominated region, $t_D$ exceeds the inflow time-scale $t_a$
and a detailed treatment of the stationary diffusion is not relevant:
$t_a<t_D$ means that the gas accretes faster than a stationary 
vertical distribution of the trapped radiation could establish. However, 
the estimation (8) gives the wright limit $F^-\ll F^+$ at $r\ll r_t$. 
Indeed, when $t_a\ll t_D$ 
we have $\Pi\approx F^+t_a$, hence $F^-/F^+\approx t_a/t_D\ll 1$. 
The radiative losses are "turned off" inside $r_t$, and 
the exact value of $F^-$ is not important
for hydrodynamical behaviour of the flow. A detailed 
calculation of $F^-$ would require 2D simulation of the radiation
diffusion in the disc with a specified vertical distribution of the viscous
energy release.

For definiteness we hereafter choose $\chi=2/\sqrt{3}$ in equation (8).
\medskip

\noindent
{\it vii) Global energy conservation}

The luminosity of the disc is related to the accretion rate by (Beloborodov
et al. 1997)
\begin{equation}
    L^-=-\frac{2\pi}{c}\int\limits_{r_s}^\infty u_t F^- rdr\approx\dot{M}c^2
    \left(1+\mu_{s}\frac{u_t^{s}}{c}\right),
\end{equation}
where index "s" refers to the inner transonic edge of the disc. 
The advection effect results in that the luminosity is less than the total
power released in the disc,
\begin{eqnarray}
\nonumber
  L^+=-\frac{2\pi}{c}\int\limits_{r_s}^\infty u_t F^+ rdr.
\end{eqnarray}
The power "swallowed" by the black hole equals $L_{adv}=L^+-L^-$.

\section{Numerical solution}

To solve numerically the disc structure equations, we choose three 
independent variables $\Omega,c_s,$ and 
$\zeta=\mu(2\nu\Sigma r\sigma^r_{~\varphi}+\dot{M}u_\varphi/2\pi)$.
From equations (3),(4) we have
\begin{equation}
 \frac{d\zeta}{dr}=\frac{F^-}{c^2}\;ru_\varphi,
\end{equation}
\begin{equation}
   \frac{P}{\Sigma^2}\frac{d\Sigma}{dr}-
   3\frac{d}{dr}\left(\frac{P}{\Sigma}\right)=\frac{2\pi rc}{\dot{M}}
   \left(F^+-F^-\right).
\end{equation}
Expressing $u_\varphi,F^\pm,P,\Sigma$ in terms of $\Omega,c_s,\zeta$ we
get two differential equations for the three unknowns. The third differential
equation is the radial equation (5).
We solve the equations (5,10,11) with external boundary conditions 
${c_s}^{out},\Omega^{out},\zeta^{out}$ at a radius $r_{out}$ such that 
$r_t\ll r_{out}\ll r_{eq}$. The values of $c_s^{out}$ and 
$\Omega^{out}=\Omega_K^+$ are taken from the relativistic version of
the standard model (Novikov \& Thorne 1973; Page \& Thorne 1974) with 
the corrected vertical balance (Riffert \& Herold 1995). 
The value of $\zeta^{out}$ is the eigen value of the problem, see below. 

We use the relaxation technique. 
As first approximation we assume $\Omega(r)=\Omega^+_K(r)$ and 
solve equations (10,11) for $c_s,\zeta$. Substituting the obtained solution to 
the radial equation (5), we calculate $\Omega^\prime(x)$ needed to fulfill 
this equation. Then we perform the relaxation step changing $\Omega(r)$ to
$\Omega(r)+\epsilon\delta\Omega(r)$ where $\delta\Omega=\Omega^\prime-\Omega$
and $\epsilon<1$ is the relaxation parameter. With new $\Omega(r)$ we solve
again equations (10,11), and so on until $\delta\Omega$ converges 
to zero at all $r$. This method is numerically unstable unless some smoothing
procedure is used to suppress numerical oscillations in $\delta\Omega(r)$.
We have found the needed procedure as a combination of two-point and
three-point smoothing, and determined $\Delta\Omega=\Omega-\Omega_K^+$ with 
an accuracy of $\sim 0.1$ \%. 

This technique allows to obtain a solution of equations (5,10,11) for
any given $\zeta^{out}$. Then we adjust $\zeta^{out}$ to fulfill the 
regularity condition at the sonic radius, $r_s$, and find the transonic 
solution. The regularity 
condition can be written as $N=D=0$ where $N$ and $D$ are the numerator
and denumerator in the radial equation with explicitly expressed derivative
of velocity, $du^r/dr=N/D$. In this way, we find $r_s$ with an accuracy of
$\sim 1$ \% 
and calculate the disc structure at $r>r_s$ (inside $r_s$ the gas is almost
in free fall).

\begin{figure*}
\begin{center}
\epsfxsize=17.5cm 
\epsfbox{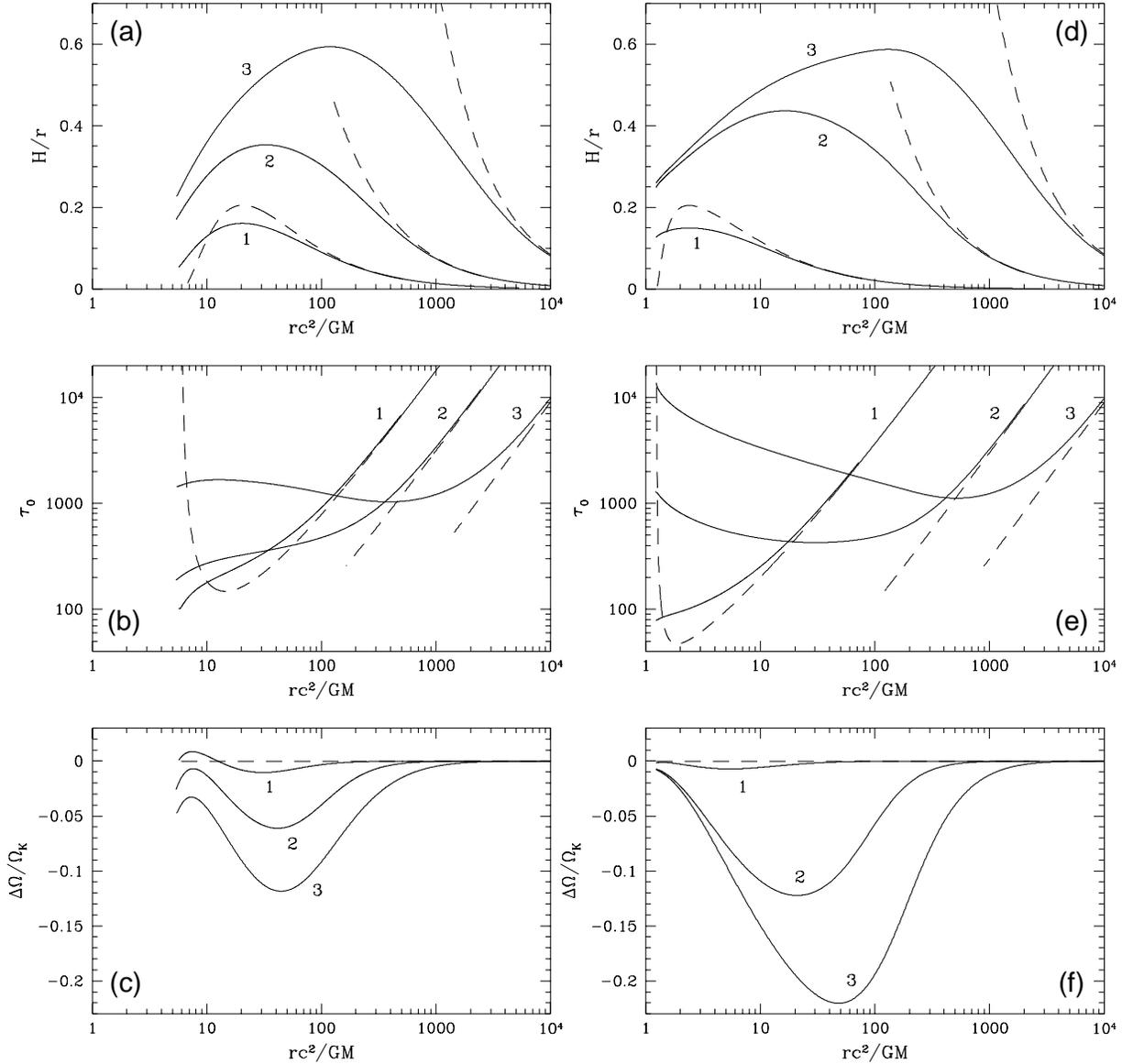}
\end{center}
\label{fig:dc}
\caption{The structure of the disc around a black hole with 
$a_*=0$ (a--c) and $a_*=0.998$ (d--f) in the case of viscosity parameter
$\alpha=0.1$. The upper panels show the relative height of the disc, 
the panels in the middle show the Thomson optical depth of the flow,
and the bottom panels show the deviation from Keplerian rotation.
All these quantities are independent of the black hole mass. 
Curves with different numbers correspond to different accretion rates:
1 -- $\dot{m}=\dot{m}_{cr}$ ($\dot{m}_{cr}=17.5$ in the case of $a_*=0$ 
and $\dot{m}_{cr}=3.11$ in the case of $a_*=0.998$), 
2 -- $\dot{m}=100$, 3 -- $\dot{m}=1000$. 
The dashed lines display the relativistic standard model.
} 
\end{figure*}

Example solutions outside $r_s$ are shown in Fig.~1 for the cases of $a_*=0$ 
and $a_*=0.998$ (the represented magnitudes are independent of the black 
hole mass). In these models $\alpha=0.1$. To a good accuracy,
solutions with different $\alpha^\prime$ can be 
obtained from the solutions with $\alpha=0.1$ using a simple scaling law:
$\Sigma^\prime=\Sigma\alpha/\alpha^\prime$, $P^\prime=P\alpha/\alpha^\prime$,
with the same $H(r)$, $\Delta\Omega(r)$, and $c_s(r)$. Essential 
deviation from this law appears only when $\alpha\rightarrow 1$. Then the
sonic radius moves outwards being beyond the radius of the Keplerian
marginally stable orbit, $r_{ms}$. At small $\alpha$, $r_s<r_{ms}$. 

The relativistic version of the standard model is shown
by the dashed lines in Fig.~1. The accretion flow deviates from this model
at the trapping radius, $r_t$. Inside $r_t$, the condition $\dot{M}c^2\mu>
2\pi r^2F^-$ takes place which means that the radial flux of internal 
energy exceeds the local radiative losses, i.e., the flow is advection 
dominated. The trapping radius is practically independent of $\alpha$,
and the dependence on $\dot{m}$ is given in Fig.~3.
From Fig.~1 one can see that: 1) the advection effect reduces the height, $H$, 
below the standard value, so that the relative disc height, $H/r$, is kept 
modest even at $\dot{m}\simgreat 10\dot{m}_{cr}$, 2) the Thomson optical depth 
of the disc, $\tau_0$, has a minimum of $\sim 100\;(\alpha/0.1)^{-1}$ at 
$\dot{m}\sim\dot{m}_{cr}$, and 3) the maximum deviation from Keplerian rotation 
is $\sim$ 20 \%.

\begin{figure}
\begin{center}
\epsfxsize=8.4cm 
\epsfbox{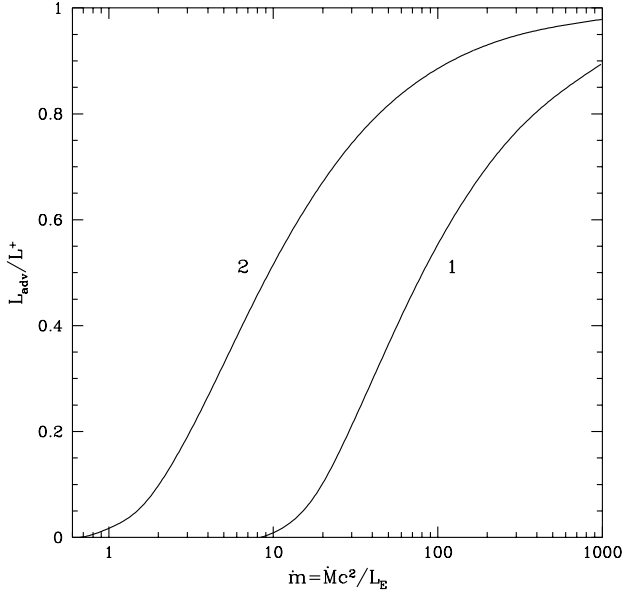}
\end{center}
\caption{The fraction of the total power released in the disc
that is eventually advected into the black hole (see item {\it vii} of 
Section 2). Curves 1 and 2 correspond to $a_*=0$ and $a_*=0.998$ 
respectively.}
\end{figure}

The critical accretion rate $\dot{m}_{cr}=17.5$ in the case of $a_*=0$
and $\dot{m}_{cr}=3.11$ in the case of $a_*=0.998$. At $\dot{m}>\dot{m}_{cr}$
a substantial fraction of the total energy released in the disc is advected 
into the black hole, see Fig. 2.

Note that in the standard disc with $\dot{m}<\dot{m}_{cr}$ the density,
$n=\Sigma/2Hm_p$, 
scales as $\dot{m}^{-2}$. 
In the advection dominated regime ($\dot{m}>\dot{m}_{cr}$, $r\ll r_t$),
the density scales as $\dot{m}$. It means that at $\dot{m}\sim\dot{m}_{cr}$
there must be a minimum of the density. At this minimum a strong overheating
of the flow is possible as discussed in next Section.

\section{Overheating in the central region}
\medskip

An accretion disc can be considered as a two-fluid flow "plasma + radiation".
In the case of a small viscosity, the disc is dense and the plasma and radiation 
are close to thermodynamic equilibrium at a common temperature 
$T\approx T_{eff}=(w/a_r)^{1/4}$ where $w=3P/2H$ is the radiation density 
calculated in Section 3, and $a_r=7.56\cdot 10^{-15}$ is the radiation 
constant. [Note that $T_{eff}$ inside the disc differs from the surface 
effective temperature $T_{eff}^s=(F^-/2\sigma)^{1/4}$ where $\sigma=a_rc/4$ 
is the Stefan-Boltzmann constant.] 
However, when $\alpha > 0.03$ and $\dot{m}$ approaches $\dot{m}_{cr}$
a strong deviation from the equilibrium occurs in the inner region 
of the disc. In this case, the accreting plasma has a low density and therefore, 
a low emission capability. As a result, the plasma does not manage to reprocess 
the released energy into Planckian radiation, and the balance between heating 
and heat-removal through radiative cooling maintains a temperature 
$T\gg T_{eff}$. Then the radiation is concentrated in the Wien peak of 
temperature $T$ and its density, $w$, is much less than the corresponding 
Planckian value, $w_{pl}=a_rT^4$, i.e., the flow is far from the black body 
state. 
\bigskip

\begin{center}
  {\it a) Radius of the overheated region}
\end{center}
\medskip

In the outer "black  body" region of the disc, the free-free emission 
capability, $\dot{w}_{ff}$, of the plasma with $T=T_{eff}$ exceeds the heating 
rate, $\dot{w}^+=F^+/2H$. The boundary $r_*$ between the black body and the 
overheated regions can be estimated from the condition
\medskip
\begin{equation}
  \dot{w}^+=\dot{w}_{ff}(T_{eff}) \quad {\rm at} \quad r=r_*, 
\end{equation}
\begin{eqnarray}
\nonumber 
  \dot{w}_{ff}=1.6\cdot 10^{-27}n^2\sqrt{T}.
\end{eqnarray}
At exact thermodynamic equilibrium, free-free emission would be
balanced by free-free absorption, and $\dot{w}_{ff}$ can be written as
$\dot{w}_{ff}=cn{\bar\sigma}_{ff}w_{pl}$. $\bar{\sigma}_{ff}$ introduced 
in this way has the meaning of an effective cross section for absorption of 
the Planckian radiation; it is $\approx 4$ times the Rosseland averaged 
$\sigma_{ff}$.

Radius $r_*$ is usually estimated from the condition that the 
effective optical depth of the flow $\tau_*=(\tau_0\tau_{ff})^{1/2}=1$
where $\tau_{ff}=Hn{\bar\sigma}_{ff}$ (Rosseland $\sigma_{ff}$ is also
often taken
instead of ${\bar\sigma}_{ff}$). For the standard disc this condition
is equivalent to equation (12). In that case, the radiation 
density inside the disc equals $w\approx\dot{w}^+t_D$ where $t_D=\tau_0H/c$ 
is the diffusion time-scale. Then equality (12) with $w\approx w_{pl}$ gives 
$\tau_*\approx 1$, and the decoupling of $w$ below $w_{pl}$ happens if 
$\tau_*<1$. The condition $\tau_*<1$ means that a deviation from 
thermodynamic equilibrium in the standard disc occurs if the bulk of 
radiation diffuses out without absorption and re-emission. 

This condition, however, is not relevant in the advection dominated regime.
As the radiation is advected rather than escape, the 
time-scale for free-free absorption should be compared 
with the inflow time-scale rather than with the diffusion time-scale.
Then the condition $\tau_*<1$ should be replaced by 
$\tau_*< (t_D/t_a)^{1/2}\sim r_t/r$ that follows from equation (12). 

\begin{figure}
\begin{center}
\epsfxsize=8.4cm 
\epsfbox{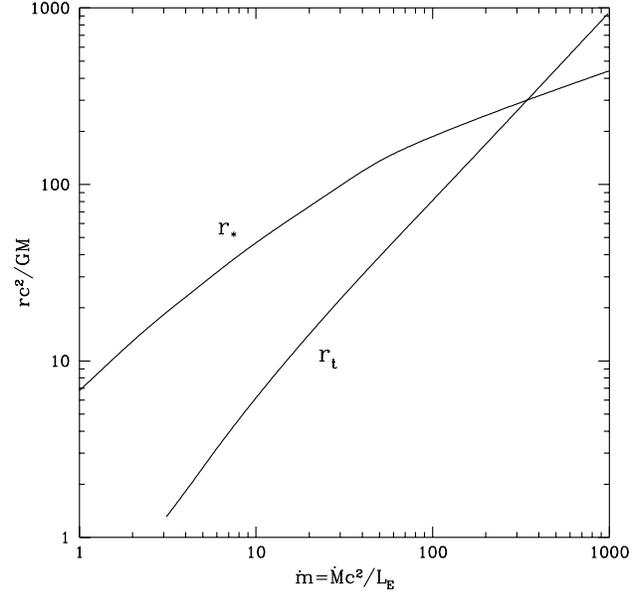}
\end{center}
\caption{The radius of the overheated region $r_*$ versus $\dot{m}$
in the case of $\alpha=0.3$, $M=10^8M_\odot$, $a_*=0.998$. 
For comparison the trapping radius, $r_t$, is also plotted.}
\end{figure}

We have calculated $r_*$ from equation (12) for the case of $\alpha=0.3$,
$M=10^8M_\odot$, $a_*=0.998$. In Fig.~3, $r_*$ is shown versus $\dot{m}$
and compared with the trapping radius, $r_t$. One can see that the entire
advection dominated region is overheated at $\dot{m}< 300$. 

\bigskip

\begin{center}
{\it b)  Cooling rate inside $r_*$}
\end{center}

The overheated plasma inside $r_*$ is cooled by Comptonized free-free
emission with rate
\begin{equation}
   \dot{w}_{cool}=A\dot{w}_{ff}, \qquad r\ll r_*,
\end{equation}
where $A(n,T)$ is an amplification factor due to Compton upscattering of 
photons emitted at low energies $h\nu<kT$. This factor is given by
(e.g. Rybicki \& Lightman 1979)
\begin{equation}
  A=1+\frac{3}{4}\ln^2 x_{coh},
\end{equation}
where $x_{coh}=h\nu_{coh}/kT$ is the dimensionless photon energy below 
which the absorption time-scale is shorter than the time-scale for shifting 
in frequency due to Comptonization. $x_{coh}$ is determined by the equation
$\sigma_{ff}(x_{coh})=8\theta\sigma_T$, where $\theta=kT/m_ec^2$, and
$\sigma_{ff}(x)=1.8\cdot 10^{-33}\ln(2.25/x)(1-e^{-x})x^{-3}\theta^{-7/2}
n\sigma_T$. Expression (14) for the Compton enhancement factor assumes that
photons with $x>x_{coh}$ upscatter to the Wien peak before they can
escape the disc or advect to the black hole.
The time-scale for upscattering to the Wien peak is
$t_C=\ln(x_{coh}^{-1})/8\theta n\sigma_T c$. Hence,
$t_C/t_D\sim \ln(x_{coh}^{-1})/y$ where $y=4\theta\tau_0^2$ is the 
Kompaneets' parameter. In the considered situation, $y\gg 1$ and photons with 
$x>x_{coh}$ do comptonize to the Wien peak before they can escape. 
Inside the trapping radius, $t_C$ should be compared with $t_a$. 
We check in the calculated models that $t_C$ is shorter than $t_a$ as well.

\begin{center}
   {\it c) Heating=cooling balance}
\end{center}

Everywhere in the disc the plasma temperature is determined by the 
heating=cooling balance $\dot{w}^+=\dot{w}_{cool}$. 
Inside $r_*$, $\dot{w}_{cool}$ is given by equation (13). Outside $r_*$,
the cooling rate is equal to a difference between free-free emission and 
absorption, being proportional to a small deviation of $w_{pl}$ from $w$: 
$\dot{w}_{cool}=\dot{w}_{ff}(1-w/w_{pl})$. The transition at $r\sim r_*$ can be 
smoothly described by the following interpolation for the heating=cooling 
balance,
\begin{equation}
  \dot{w}^+=\dot{w}_{ff}\left[\frac{w}{w_{pl}}+
            A\left(1-\frac{w}{w_{pl}}\right)\right]
            \left(1-\frac{w}{w_{pl}}\right).
\end{equation}
In the limit $w\approx w_{pl}$ this equation yields 
$\dot{w}^+=\dot{w}_{ff}(1-w/w_{pl})$ while in the limit $w\ll w_{pl}$ 
it transforms into $\dot{w}^+=A\dot{w}_{ff}$.

\begin{center}
   {\it d)  Results}
\end{center}

\begin{figure}
\begin{center}
\epsfxsize=8.4cm 
\epsfbox{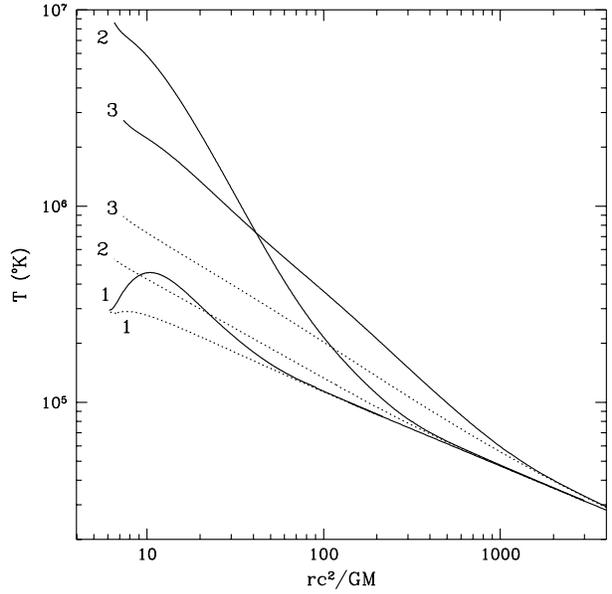}
\end{center}
\caption{ The temperature of the disc with $\alpha=0.3$ in the case of 
$a_*=0$:
1 -- $\dot{m}=10$, 2 -- $\dot{m}=100$, 3 -- $\dot{m}=1000$. The dotted 
lines show the effective temperature, $T_{eff}$, of the radiation field
in the disc.}
\end{figure}

\begin{figure}
\begin{center}
\epsfxsize=8.4cm 
\epsfbox{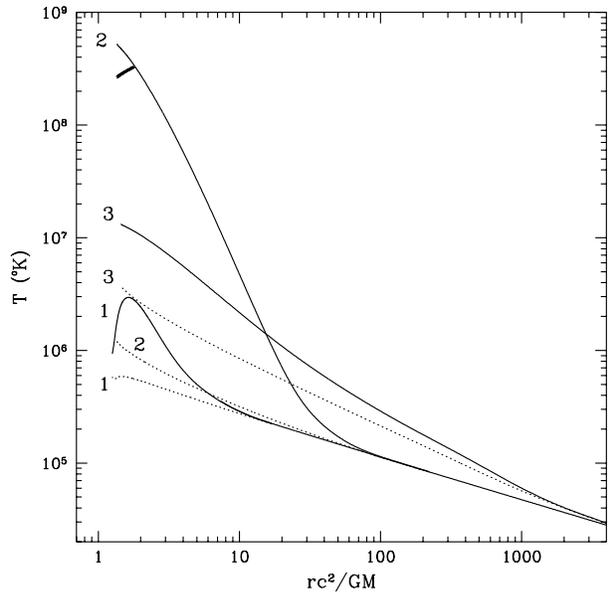}
\end{center}
\caption{The temperature of the disc with $\alpha=0.3$ in the case of 
$a_*=0.998$: 1 -- $\dot{m}=1$, 2 -- $\dot{m}=10$, 3 -- $\dot{m}=1000$. 
The dotted lines display the effective temperature, $T_{eff}$, of the 
radiation in the disc. The heavy line shows the temperature
for which $\nu_B=\nu_{coh}$ at $B=B_{eq}$ (see the text).}  
\end{figure}

\begin{figure}
\begin{center}
\epsfxsize=8.4cm 
\epsfbox{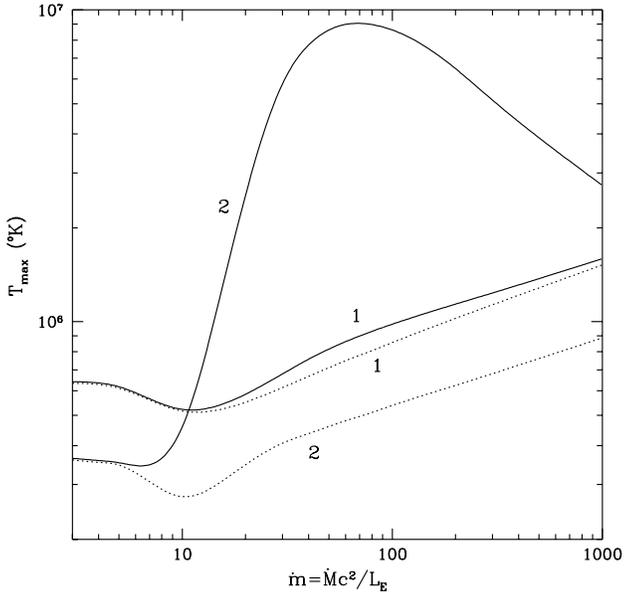}
\end{center}
\caption{The maximum temperature in the disc as a function of the 
accretion rate for the case of $a_*=0$: 1 -- $\alpha=0.03$, 2 -- 
$\alpha=0.3$. $T_{eff}$ at the maximum is displayed by the dotted lines.}
\end{figure}

\begin{figure}
\begin{center}
\epsfxsize=8.4cm 
\epsfbox{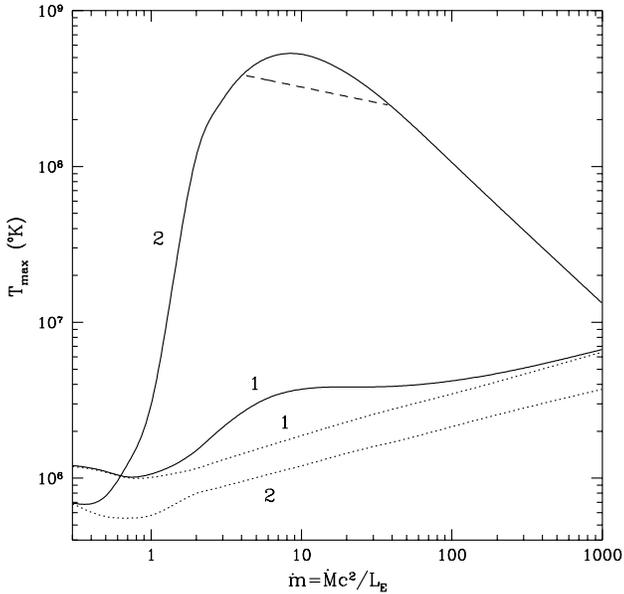}
\end{center}
\caption{The maximum temperature in the disc as a function of the 
accretion rate for the case of $a_*=0.998$: 1 -- $\alpha=0.03$, 
2 -- $\alpha=0.3$. $T_{eff}$ at the maximum is displayed by the dotted lines.
Above the dashed line, $\nu_B$ exceeds $\nu_{coh}$ in the inner hottest 
region of the disc.}
\end{figure}

Using the balance (15), we have calculated the temperature in the disc 
around a massive black hole $M=10^8M_\odot$ with spin $a_*=0$ and
$a_*=0.998$. 
The results are shown in Fig.~4,5 for the case of $\alpha=0.3$.
The strong overheating, $T\gg T_{eff}$, occurs inside the radius $r_*$.
The maximum temperature is reached in the innermost region. 
It is $\sim 3\cdot 10^8$ K in the case of a rapidly rotating black hole 
and $\sim 10^7$ K in the case of a Schwarzschild black hole. 
In Fig.~6,7 we show the dependence of the maximum temperature on the 
accretion rate in the cases of $\alpha=0.03$ and $\alpha=0.3$.

A question of interest is whether the protons can be thermally decoupled from 
the electrons at such temperatures. The time-scale for Coulomb energy exchange
between protons of temperature $T_p$ and electrons of temperature $T_e$
is given by (Landau \& Lifshitz 1981)
\begin{eqnarray}
\nonumber
   t_{ep}=\sqrt{\frac{\pi}{2}}\frac{m_p}{m_e}
          \left(\frac{kT_e}{m_ec^2}\right)^{3/2}
          \frac{1}{\ln\Lambda\;n\sigma_Tc}\approx 12.5 \frac{T_e^{3/2}}{n},
\end{eqnarray}
where $\ln\Lambda\approx 20$ is the Coulomb logarithm. Even in the most hot 
models with $\alpha=0.3$ and $a_*=0.998$, $t_{ep}\ll t_a$ and the accreting
plasma can be well described in the one-temperature approximation 
$T_e\approx T_p\approx T$. 

Then we have compared the thermal plasma pressure, $p_{gas}=2nkT$, with 
the radiation pressure, $p_{rad}=w/3$. Even in the hottest region of
the disc with $\alpha=0.3$,
the plasma pressure does not exceed $\sim 10^{-3} p_{rad}$.
So, the calculated radiation dominated models are self-consistent.
Note, that accretion flows with $\alpha\rightarrow 1$ 
are very unlikely (it would imply a turbulence of scale 
$H$ with sound speed).
Anyway, such flows could not be much hotter because at higher temperatures
additional cooling mechanisms become important: Comptonization of 
cyclotron radiation and collective waves in the plasma, 
as noted by Shakura \& Sunyaev (1973).
Cyclotron radiation and collective waves are additional sources of soft 
photons which can be upscattered to the Wien peak and cool the plasma efficiently. 
Roughly, the cyclotron source starts to be important when the gyrofrequency,
$\nu_B=eB/2\pi m_ec$, exceeds $\nu_{coh}$ (a more detailed treatment including
higher cyclotron harmonics is given in Gnedin \& Sunyaev 1973). In a similar
way, Comptonization of collective waves can be important when $\nu_{pl}
\simgreat\nu_{coh}$, where $\nu_{pl}=(ne^2/\pi m_e)^{1/2}$ is the plasma 
frequency. In case the magnetic field is comparable to the equipartition value 
$B_{eq}=(8\pi w)^{1/2}$, the gyrofrequency exceeds $\nu_{coh}$ in the disc 
sooner than the plasma frequency does. 
In Fig.~5, the heavy line shows the temperature at which the 
gyrofrequency equals $\nu_{coh}$. The corresponding 
maximum temperature in the disc is shown by the dashed line in Fig.~7.

\section{Conclusions}

We have investigated the relativistic accretion disc around a Kerr 
black hole with a super-Eddington accretion rate. The disc has a modest
relative height, $H/r < 0.4$, up to $\dot{m}\sim 10\dot{m}_{cr}$, and a 
luminosity near the Eddington limit. The bulk of the energy released in the
inner region of the  
disc is advected into the black hole. The angular velocity of rotation
deviates from Keplerian up to 20 \%. We paid particular attention to 
the case of a large viscosity parameter $\alpha > 0.03$. 
In this case, the accretion flow 
deviates from thermodynamic equilibrium and overheats in the central region.
The hottest flow has accretion rate $\dot{m}\sim 3\dot{m}_{cr}$.
We have calculated the maximum temperature in the disc around a massive 
black hole, $M=10^8M_\odot$, with $\alpha=0.3$. 
For a Schwarzschild black hole it equals $\sim 10^7$ K.
For a rapidly rotating black hole, the maximum 
temperature is of order $\sim 3\cdot 10^8$ K. It far exceeds
the effective temperature corresponding to thermodynamic equilibrium. 
However, it is not large enough for thermal decoupling of the protons and 
transition to the ion pressure dominated regime of accretion.

The employed vertically integrated model gives approximate 
characteristics of the bulk of the gas flowing in the disc, and does not
describe physical conditions in the upper layers (in the "skin" of the
disc) where the spectrum of emerging radiation is formed. 
To evaluate the spectrum
of the super-Eddington disc a more detailed treatment of the vertical energy
transfer is needed which is likely to demand a 2D simulation of the advective
region. Besides, the 
disc spectrum may be strongly affected by a corona activity above the surface.
It is clear, however, that the surface of the flow with large $\alpha$ and
a large accretion rate must be much hotter than the effective surface 
temperature, $T^s_{eff}=(F^-/2\sigma)^{1/4}$, and an activity of the corona 
may make the spectrum only harder. 
X-ray emission of the innermost region of the 
super-Eddington disc may help to reconcile the observed quasar spectra 
with accretion disc models.

\section*{Acknowledgments}

I am grateful to M.A. Abramowicz, I.V. Igumenshchev, and I.D. Novikov
for discussions. I thank the Theoretical Astrophysics Center  
for hospitality and acknowledge partial support by RFFI grant 97-02-16975.


\end{document}